\documentclass[journal=aelccp,manuscript=letter]{achemso}

\usepackage[version=3]{mhchem} 

\author{Vlad~Alexandru~Dragomir}
\affiliation[UdeM]{D\'epartement de physique, Universit\'e de Montr\'eal, C.P.\ 6128, Succursale centre-ville, Montr\'eal (Qu\'ebec) H3C~3J7, Canada}
\author{Stefanie~Neutzner}
\affiliation[IIT]{Center for Nano Science and Technology@PoliMi, Istituto Italiano di Tecnologia, via Giovanni Pascoli 70/3, 20133 Milano, Italy}
\author{Claudio~Quarti}
\affiliation[Mons]{Laboratory for Chemistry of Novel Materials, Department of Chemistry, Universit\'e de Mons, Place du Parc 20, 7000, Mons, Belgium}
\author{Daniele~Cortecchia}
\affiliation[IIT]{Center for Nano Science and Technology@PoliMi, Istituto Italiano di Tecnologia, via Giovanni Pascoli 70/3, 20133 Milano, Italy}
\author{Annamaria~Petrozza}
\affiliation[IIT]{Center for Nano Science and Technology@PoliMi, Istituto Italiano di Tecnologia, via Giovanni Pascoli 70/3, 20133 Milano, Italy}
\author{Sjoerd~Roorda}
\affiliation[UdeM]{D\'epartement de physique, Universit\'e de Montr\'eal, C.P.\ 6128, Succursale centre-ville, Montr\'eal (Qu\'ebec) H3C~3J7, Canada}
\author{David~Beljonne}
\affiliation[Mons]{Laboratory for Chemistry of Novel Materials, Department of Chemistry, Universit\'e de Mons, Place du Parc 20, 7000, Mons, Belgium}
\author{Richard~Leonelli}
\affiliation[UdeM]{D\'epartement de physique, Universit\'e de Montr\'eal, C.P.\ 6128, Succursale centre-ville, Montr\'eal (Qu\'ebec) H3C~3J7, Canada}
\author{Ajay~Ram~Srimath~Kandada}
\affiliation[IIT]{Center for Nano Science and Technology@PoliMi, Istituto Italiano di Tecnologia, via Giovanni Pascoli 70/3, 20133 Milano, Italy}
\alsoaffiliation[GTchem]{School of Chemistry and Biochemistry, Georgia Institute of Technology, 901 Atlantic Drive, Atlanta, Georgia 30332, USA}
\alsoaffiliation[GTphys]{School of Physics, Georgia Institute of Technology, 837 State Street NW, Atlanta, Georgia 30332, USA}
\email{srinivasa.srimath@iit.it}
\author{Carlos~Silva}
\affiliation[GTchem]{School of Chemistry and Biochemistry, Georgia Institute of Technology, 901 Atlantic Drive, Atlanta, Georgia 30332, USA}
\alsoaffiliation[GTphys]{School of Physics, Georgia Institute of Technology, 837 State Street NW, Atlanta, Georgia 30332, USA}
\email{carlos.silva@gatech.edu}

\title[Lattice dynamics of hybrid 2D lead-halide perovskites]
  {Lattice vibrations and dynamic disorder in two-dimensional hybrid lead-halide perovskites}

\begin{document}







\newpage
\begin{abstract}
  By means of non-resonant Raman spectroscopy and density functional theory calculations, we measure and assign the vibrational spectrum of two distinct two-dimensional lead-iodide perovskite derivatives. These two samples are selected in order to probe the effects of the organic cation on lattice dynamics. One templating cation is composed of a phenyl-substituted ammonium derivative, while the other contains a linear alkyl group. We find that modes that directly involve the organic cation are more prevalent in the phenyl-substituted derivative. Comparison of the temperature dependence of the Raman spectra reveals differences in the nature of dynamic disorder, with a strong dependence on the molecular nature of the organic moiety.
\end{abstract}

\newpage
Hybrid organic-inorganic metal-halide perovskites (HOIPs) are a class of ionic semiconductors that are characterized by the `softness' of the crystal lattice
~\cite{guo_interplay_2017}. This gives rise to complex lattice dynamics that define the nature of charge transport~\cite{milot_charge-carrier_2016, tsai_design_2018} and relaxation processes~\cite{straus_direct_2016}, as well as the nature of dynamic disorder~\cite{kang_dynamic_2017}. Although it was initially argued that the motion induced by the organic-inorganic interactions has considerable impact on the dynamic disorder of the lattice~\cite{baranowski_static_2018, brivio_lattice_2015}, later works demonstrated that such motion is ubiquitous to the metal-halide perovskite structure, even within all-inorganic frameworks~\cite{yaffe_local_2017}.  However, some experiments have also provided evidences for the non-trivial influence of the organic cation motion on the carrier induced lattice deformation and subsequently on the polaron formation dynamics~\cite{miyata2017}.

\begin{figure}[th]
  \centering
    \includegraphics[width=0.65\textwidth]{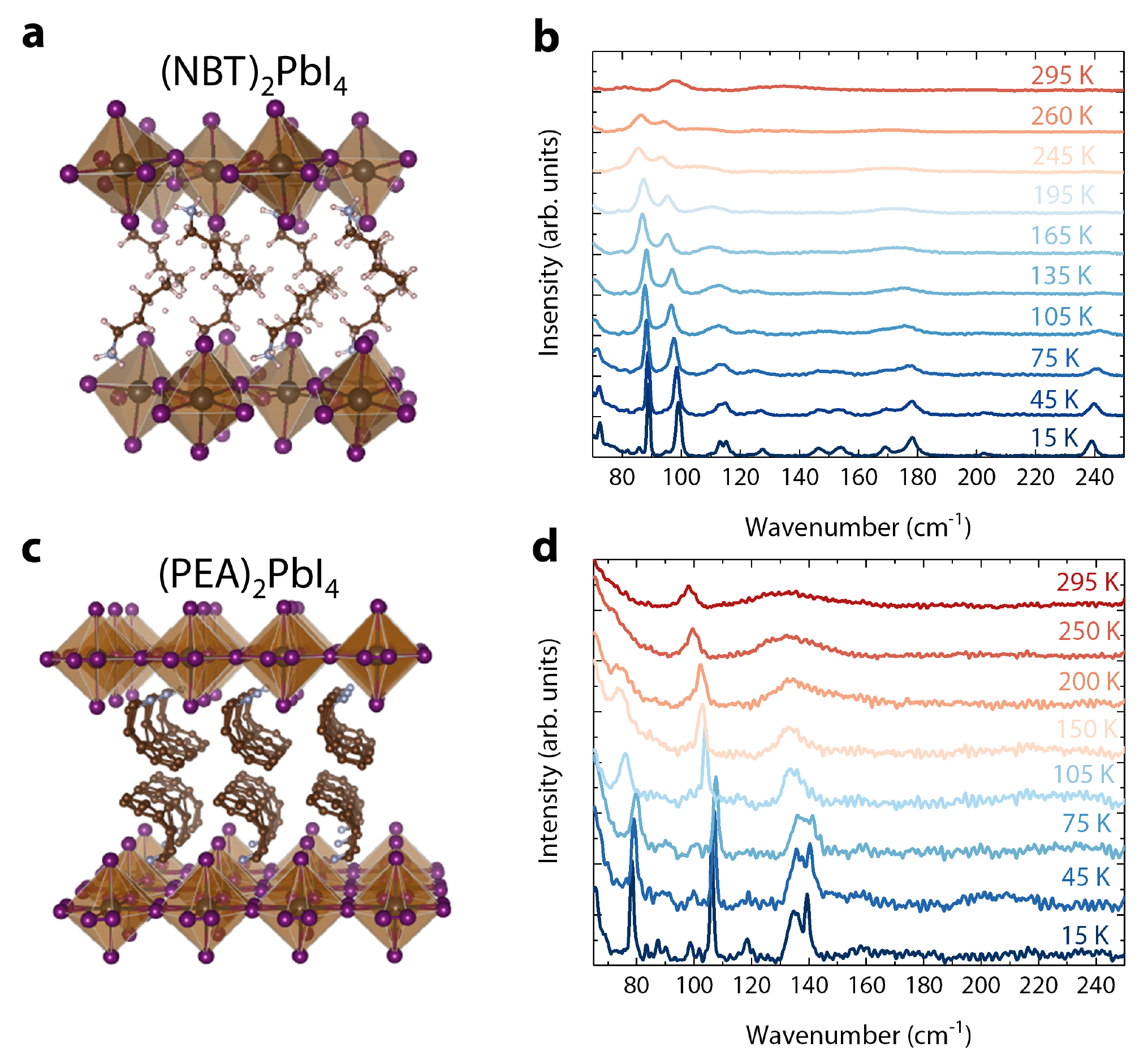}
\caption{Hybrid organic-inorganic perovskites (HOIP's). (a) Representation of the (NBT)$_{2}$PbI$_{4}$ perovskite.(b) Raman spectra of (NBT)$_{2}$PbI$_{4}$ at 15 K. (c) Representation of the (PEA)$_{2}$PbI$_{4}$ perovskite. The monolayered  PbI$_{4}$ octahedra are separated organic cations. (d) Raman spectra of (PEA)$_{2}$PbI$_{4}$ at 15 K.} 
\label{figure1}
\end{figure}
We have recently demonstrated that such polaronic effects can be perceived even within two-dimensional counterparts of the 3D HOIPs. When chosen appropriately, the organic cation in the crystalline motif in perovskitoid derivatives 
separates the single layers of metal-halide octahedra lattices planes 
and thus electronically decouples them from each other. This results in quantum-well-like structures as exemplified in Fig.~\ref{figure1}(a) and (b) for two distinct cations, n-butly ammonium (NBT) and phenlyelthylammonium (PEA) respectively. Strong quantum and dielectric confinement effects give rise to Wannier excitons with extraordinarily high binding energy compared to III-V semiconductor quantum wells, for example, in the order of 200--300\,meV~\cite{blancon_scaling_2018}. 
However, as opposed to Wannier excitons in quantum wells, the exciton absorption lineshape exhibits rich spectral structure that we have attributed to polaronic effects~\cite{neutzner_exciton-polaron_2018,felixthouin_phonon_2018} --- excitons induce long-range lattice deformations due to Coulomb interactions. The fundamentally important question arising from this identification is the following: how do the lattice dynamics imposed by the cation A influence the excitonic properties of this class of materials? 

In this article, we carry out non-resonant Raman spectroscopy on two derivatives of a \emph{single layer} lead-iodide perovskite, \ce{A2PbI4} (A = n-butylammonium and phenylethylammonium), as a function of temperature over the range 15--295\,K. In one sample, the organic cation A consists of an ammonium derivative containing a linear alkyl chain, while in the other one it includes phenyl substitution. We find that the temperature-dependent Raman spectrum is strongly dependent on the nature of the cation, with the organic cation contributing more directly to Raman-active modes with frequency $\leq 200$\,cm$^{-1}$. 
We discuss the evolution of the Raman spectra with temperature in light of these differences between the two samples investigated, and we conclude that the nature of dynamic disorder at room temperature is distinct.

\begin{figure}[tbh]
\centering
\includegraphics[width=0.65\textwidth]{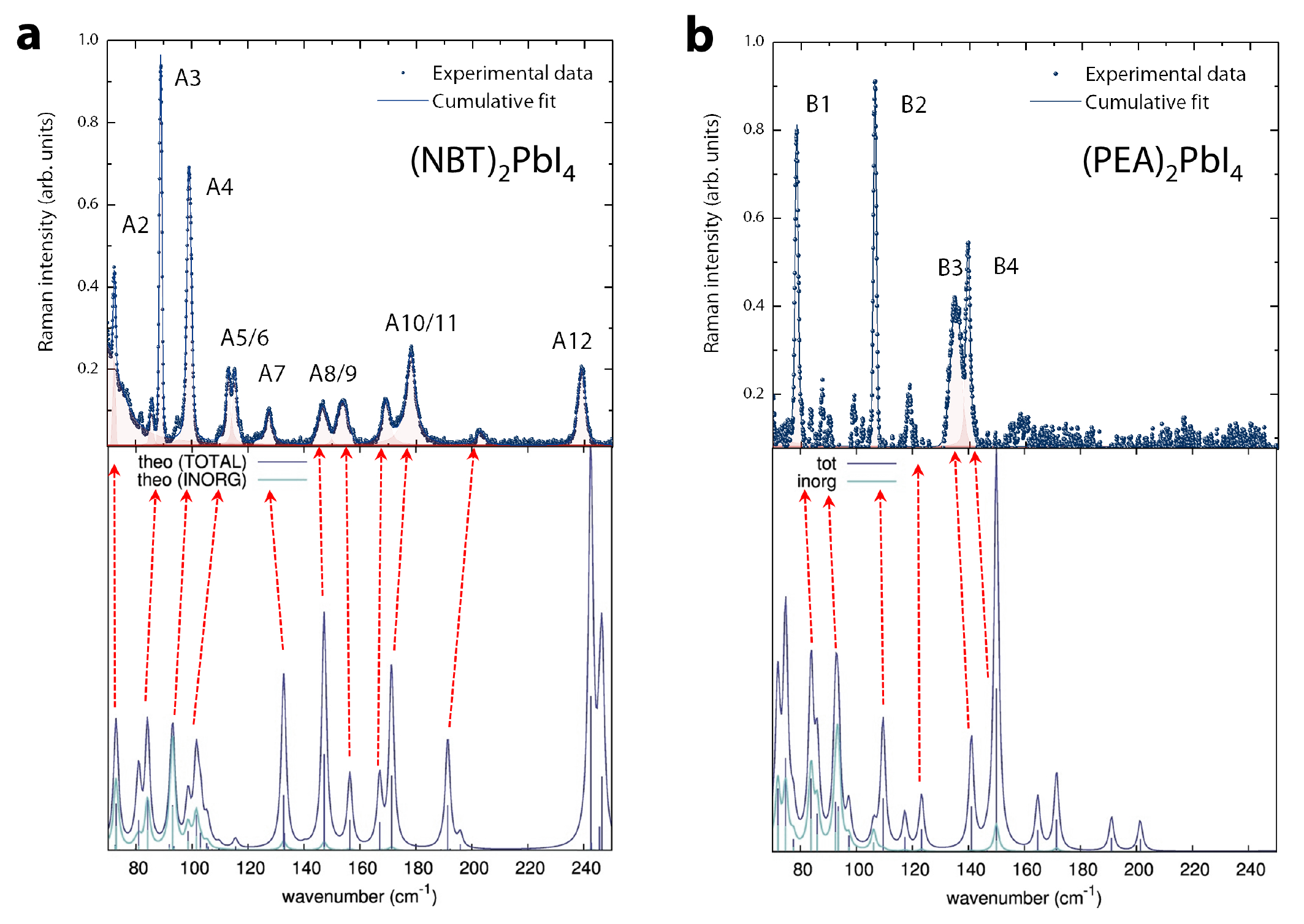}
\caption{Raman Stokes spectra of (NBT)$_{2}$PbI$_{4}$ and (PEA)$_{2}$PbI$_{4}$. (a) Experimental and theoretical assignment of Raman Stokes spectra of single crystals of (NBT)$_{2}$PbI$_{4}$ at 15\,K . (b) Experimental and theoretical assignment of Raman Stokes spectra of single crystals of (PEA)$_{2}$PbI$_{4}$ at 15 K. The measurements were done in vacuum using a cold finger cryostat at $730$\,nm with a pump power of $100$\,mW.}
\label{figure2}
\end{figure}

\section{Results}

In the current work, we are particularly interested in discerning the evolution of the vibrational spectra below room temperature. \ce{(NBT)2PbI4} undergoes a crystal phase transition in the range of 220 - 260\,K (depending on the direction of the temperature scan), below which the \ce{PbI6} octahedra are subjected to significant out-of-plane tilt, as represented in Fig.~\ref{figure1}(a). \ce{(PEA)2PbI4}, on the other hand does not show any first order phase transition in the probed temperature range, as confirmed by X-ray scattering measurements at least till about 77\,K~\cite{thouin_stable_2018}. The corresponding crystal structure with minimum octahedral distortion is shown in Fig.~\ref{figure1}(c). Fig.~\ref{figure1}(b) and (d) show the non-resonant Raman spectra taken from single crystals of \ce{(NBT)2PbI4} and \ce{(PEA)2PbI4} respectively over a wide temperature range.    

Close to room temperature, the Raman response of both the samples are composed of broad spectral lineshapes with peaks centered around 100 \,cm$^{-1}$ and 130\,cm$^{-1}$. A broad peak leading to a central peak is also evident, especially in the case of \ce{(PEA)2PbI4}. Yaffe~et~al.~\cite{yaffe_local_2017} made similar observations from temperature dependent low-frequency Raman measurements on hybrid lead-halide perovskite crystals. They observed that the Raman spectra are diffuse with continuum underlying broad Raman transitions at the positions corresponding to the softened modes. By performing molecular dynamics simulations, they have found that the structural fluctuations are prevalent in halide perovskites and they also identify the vibrational motifs that have a large amplitude at higher temperatures.  As elaborated by Yaffe et al~\cite{yaffe_local_2017}, this is an evidence of dynamic disorder within the perovskite lattice. 

Upon lowering the temperature, sharp and distinct features are observed in both the compounds. The low energy modes in the vibrational spectra of perovskite compounds correspond to the motion within the lead halide network. In spite of the fact that the investigated compounds are both based on lead iodide networks, they exhibit very distinct responses, highlighting the role played by the organic cation on the vibrational degrees of freedom. As we will discuss below, this arises not only due to the different degree of distortions induced by the organic cation within the lead iodide octahedra depending on the binding site, but also due to the relative motion introduced by the local vibrational modes of the organic ligand which changes with the chemical nature. We use the high resolution spectra obtained at 15\,K in conjunction with the density functional theory (DFT) calculations to firstly assign the observed modes to specific lattice motion and then to analyze the evolution of the each of the normal modes with temperature.

 \begin{table*}

   \caption{\ Assignment of Raman modes for (NBT)$_2$PbI$_4$} 
   \begin{tabular*}{\textwidth}{@{\extracolsep{\fill}}llllll}
   \hline
     Mode & exp. shift\,(cm$^{-1}$) & theo. shift\,(cm$^{-1}$) & Symmetry  & Description\\
     \hline
      & & & & & \\
      
       A2 & 72.32$\pm$0.07 & 73 & A$_g$  & in-phase Pb-I (apical)  \\
       & & & & stretching + octahedral rotation \\
       A3 & 88.9$\pm$0.1 & 84 & A$_g$  & in-phase Pb-I (apical)  \\
       & & & & stretching + octahedral rotation \\
       A4 & 99.63$\pm$0.1 & 93 & A$_g$  & out of phase Pb-I (equatorial)  \\
       & & & & stretching \\
       A5/6 & 114$\pm$1 & 102 & B$_{1g}$  & in-phase Pb-I (equatorial) \\
       & & & & stretching + PbI$_4$ breathing \\
       A7 & 127.5$\pm$0.5 & 133 & B$_{3g}$  & organic torsional \\
       A8/9 & 146.5$\pm$0.7/154$\pm$1 & 147/156 & A$_g$ / B$_{1g}$  & organic torsional \\
       A10/11 & 169$\pm$0.5/178.2$\pm$0.3 & 167/171 & B$_{3g}$ / A$_{g}$  & organic torsional \\
       A12 & 239.1$\pm$0.4 & 243 & A$_{g}$  & torsional of H-N-C-C angle:  \\
       & & & & involved H-bonding\\
    \hline
   \end{tabular*}
   \label{table:15K_NBT_raman_modes}
 \end{table*}

 \begin{table*}

   \caption{\ Assignment of Raman modes for (PEA)$_2$PbI$_4$}
   \begin{tabular*}{\textwidth}{@{\extracolsep{\fill}}llllll}
   \hline
   
   Mode & exp. shift\,(cm$^{-1}$) & theo. shift\,(cm$^{-1}$) & Symmetry  & Description\\
     \hline
      & & & & & \\
      
       B1 & 78.3$\pm$0.1 & 84 & A$_g$  & in-phase Pb-I stretching +   \\
        & & & & octahedral rotation \\
       B2 & 106.2$\pm$0.1 & 110 & A$_g$  & $\pi-\pi$ vibration of PEA  \\
       B3 & 134.6$\pm$0.4 & 141 & A$_g$  & $\pi-\pi$ vibration of PEA  \\
       B4/5 & 139.3$\pm$0.2 & 150 & A$_g$  & $\pi-\pi$ vibration of PEA  \\

      & & & & & \\
    \hline
   \end{tabular*}
   \label{table:15K_PEA_raman_modes}
 \end{table*}

The experimental Raman spectra of \ce{(NBT)2PbI4} and \ce{(PEA)2PbI4} taken at 15\,K are displayed in the top panels of Fig.~\ref{figure2}(a) and (b) respectively and the corresponding theoretical spectra are shown in the bottom panels. The DFT calculations are based on harmonic approximation and relies on crystallographic models of the considered compounds obtained from available X-ray diffraction data. The calculations are performed with periodic boundary conditions and localised atomic basis set as implemented in CRYSTAL17 program (more details are provided in the Methods section). As it is evident from Fig.~\ref{figure2}, the theoretical predictions match with the experimental spectral with an acceptable level of accuracy. A comparison of the experimental energies and theoretical predictions along with the associated symmetry and assignments are listed in Tables.~\ref{table:15K_NBT_raman_modes} and \ref{table:15K_PEA_raman_modes}.

In both the compounds, modes at frequencies $\leq 100$\,cm$^{-1}$, A2-A6 in \ce{(NBT)2PbI4} and B1 in \ce{(PEA)2PbI4} can be assigned to motion of the lead iodide network. This includes in phase stretching of the Pb-I bond along with rotation of the octahedra along various pseudocubic axis within the inorganic layer. We recently demonstrated existance of similar modes even below 50\,cm$^{-1}$ via impulsive stimulated Raman scattering~\cite{felixthouin_phonon_2018} and not accessible in the current experimental configuration. The appearance of multiple modes in the case of \ce{(NBT)2PbI4} with respect to \ce{(PEA)2PbI4} can be attributed to the stiffening of the lattice due to the octahedral distortion and thus increase in the energy of the vibrational modes, resulting in their appearance in the probed spectral range.      

Intriguingly, modes above 100\,cm$^{-1}$ have substantial, if not exclusive contribution from the motion of the organic cations. In the case of \ce{(NBT)2PbI4}, these correspond to various torsional motion of the C-C-C bond within the NBT molecule along with associated motion within the inorganic lattice induced via hydrogen bonding. The latter is particularly relevant for mode A12 which is associated to the torsion of the H-N-C-C bond angle and thus involves substantial motion of the amine group that binds to the inorganic framework. In the case of \ce{(PEA)2PbI4}, the three modes observed above 100\,cm$^{-1}$, B2-B4 correspond to the motion of the phenly rings that are $\pi$-stacked within the lattice.

\begin{figure}[tbh]
\centering
\includegraphics[width=0.75\textwidth]{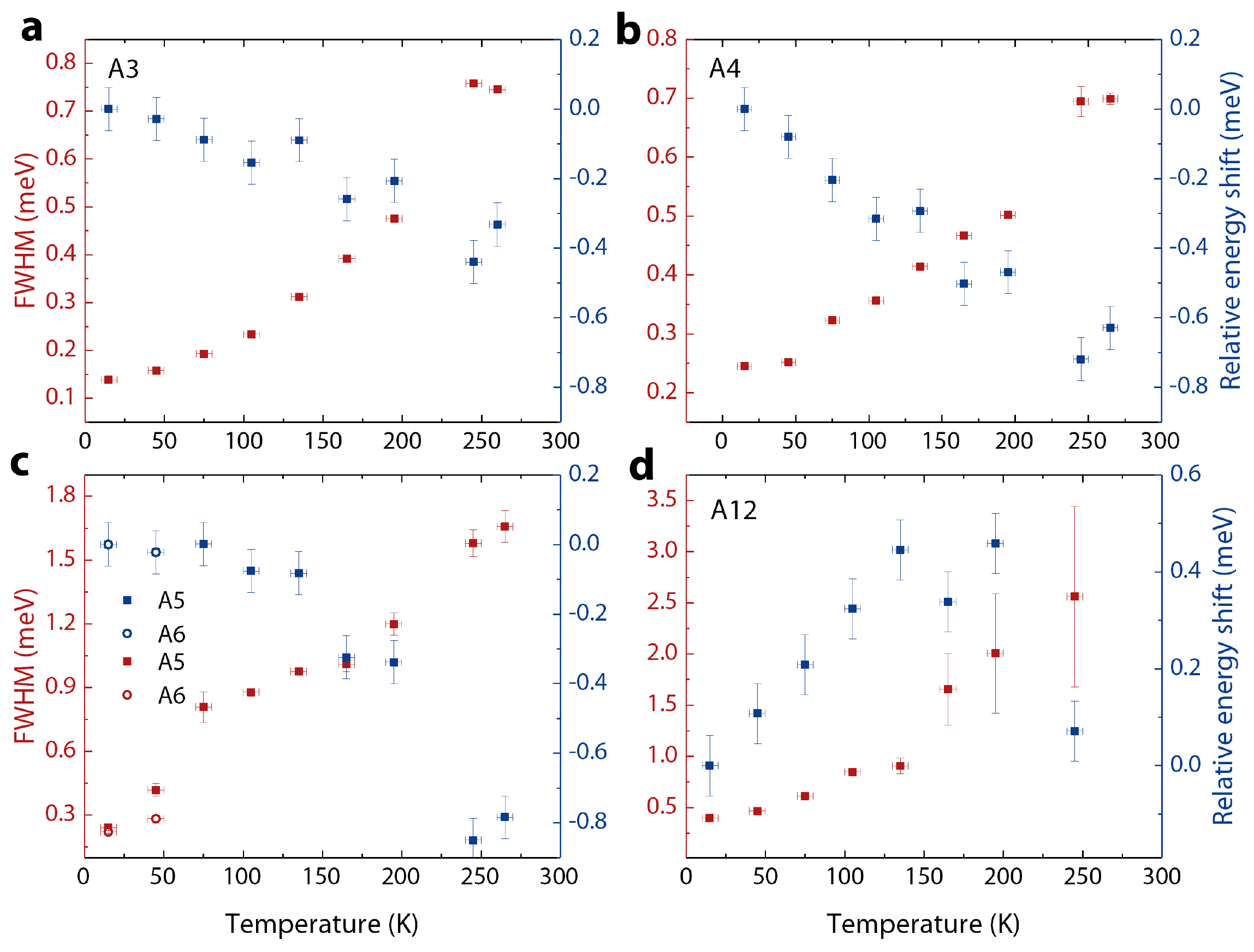}
\caption{Peak energy vs full width at half maximum (FWHM) as a function of temperature for the main modes of (NBT)$_{2}$PbI$_{4}$ at 15\,K. (a) Mode A2. (b) Mode A4. (c) A5 and A6. (d) A12 mode.}
\label{figure3}
\end{figure}
\section{Discussion}

Having successfully assigned the low temperature Raman spectra to specific lattice normal modes, we now discuss the temperature trends of the spectral energies and linewidths. Plotted in Fig.~\ref{figure3} are the relative energy shifts (with respect the 15\,K values) and the full width at half maximum of four representative modes of \ce{(NBT)2PbI4}. The choice of the modes is not only based on the existence of clear peaks over a wide temperature range, but also on their distinct nature. Modes A3 and A4 that correspond to the apical and equatorial stretching of the Pb-I bond respectively exhibit similar trends with temperature (Figs.~\ref{figure3}(a) and (b)). The mode energies monotonically reduce due to the softening of the lattice at higher temperatures. 
While this can be attributed to the thermally activated lattice expansion, Cortecchia et al~\cite{cortecchia2018structure} recently demonstrated using comprehensive X-ray diffraction data the lattice expansion is not mediated by the octahedral expansion. It is rather assisted by the tilting of the octahedra, which qualitatively explains the higher sensitivity of the equatorial stretch (A4) to temperature in comparison to the apical stretch (A3). It is intriguing here to note the anomalous trend of mode A12 whose energy increases with increasing temperature. As discussed previously, this mode corresponds to the motion of the organic cation and also involves hydrogen bonding with the inorganic lattice. Tilting of the octahedra and NBT molecule may stiffen such a motion leading to the observed increase. Another consideration which has to be quantiatively formulated pertains to the anharmonicity induced by phonon-phonon interactions at higher temperature. This may be particularly relevant to rationalize the trend of A5 shown in Fig.~\ref{figure3}(c), which exhibits a distinct behaviour above 150\,K with a rapid increase in the mode energy as well as substantial broadening of the linewidth. A quantitative analysis of these trends will provide the essential insights into the distinct trends of various modes shown in Fig.~\ref{figure3}.

Coming to the case of \ce{(PEA)2PbI4}, we reported the distinct jumps at around 100 K for (PEA)$_{2}$PbI$_{4}$, both in the Raman and linear spectra in our earlier work~\cite{thouin_stable_2018}. This is accompanied by substantial broadening of one particular mode (B3) along with suppression of another (B4) above 100\,K, which we identify to be from the organic motion. 
Given that these vibrations correspond to the $\pi$-$\pi$ motion of the phenly groups, this peculiar temperature trend can be interpreted as the increase in the disorder in the orientation of the organic cations and the intermolecular interactions, which subsequently increases both static and dynamic disorder within the lattice. Possibly, the phenyl ring stacking is driving lattice ordering and greater coupling of the seemingly localized vibrations with the lattice. As also hypothesized earlier by Thouin et al~\cite{thouin_stable_2018}, this implies an order to disorder transition at 100\,K, that concomitantly leads to a renormalization of the excitonic energies.The existence of such intermolecular interactions induced by $\pi - \pi$ stacking is peculiar to perovskite lattices coordinated by phenly based organic cations. Such disordering effects have been identified via NMR studies by Ueda et al~\cite{ueda_13c_1996}, albeit at much higher temperatures, while our observations suggest a persistance of such effects even at lower temperatures. Intriguingly, Kamminga et al~\cite{kamminga_out--plane_2018} noted in the case of \ce{(PEA)2MnCl4} that $\pi - \pi$ interactions of the phenly groups may provide an order parameter for a polar transition in such metal halide hybrids and thus act as an unique handle to control the lattice characteristics.
This shows that the lattice dynamics are intricately controlled not only by the binding amine group, the position of its binding and the length of the organic ligands, but also by the various interactions which may exist within the organic layer itself. In addition, we have previously identified and order-disorder transition for (PEA)$_{2}$PbI$_{4}$, while we do not see it in the (NBT)$_{2}$PbI$_{4}$ case. Due to the stronger intermolecular interactions in phenyl perovskites as shown by our assignment of modes in table~\ref{table:15K_PEA_raman_modes} for (PEA)$_{2}$PbI$_{4}$, we see that the dynamic disorder for such crystals a room temperature (RT) is different than perovskites made of linear alkyl chain as in the (NBT)$_{2}$PbI$_{4}$ case. This observation comes from our room temperature spectra of (NBT)$_{2}$PbI$_{4}$ and (PEA)$_{2}$PbI$_{4}$ at Fig.~\ref{figure1}, which, as we can see, are very different. Although we recognize that our study covers a limited range of frequencies, our assessment of the lattice motion in these two prototypical 2D perovskites allows us to show the importance of the organic cation substitution in these materials.

In summary we demonstrated that 
both inorganic and organic vibrational modes play an important role in lattice reorganization and dynamic disorder in 2D hybrid organic-inorganic perovskites. By means of low-frequency temperature dependent Raman spectroscopy and density functional theory calculations, we have identified distinct and narrow vibrational modes at low temperature in \ce{(NBT)2PbI4} and \ce{(PEA)PbI4}. The effect of the phenyl ligands, linear alkyl chains and their substantial difference in their dynamic disorder must be taken into account in order to have a better control over optoelectronic properties of 2D HOIPs.

\section{Experimental methods}
\subsection{Sample preparation}
Single crystals of \ce{(NBT)2PbI4} crystals are obtained via slow cooling. 100.52 mg of (NBT)I and 115.12 mg of PbI2 are dissolved in hot concentrated HI water solution (1.2 ml at $100\,^{\circ}{\rm C}$). The vial containing the solution is slowly cooled in steps of $2\,^{\circ}{\rm C}$  per hour at room temperature and left for one week in a refrigerator at $2\,^{\circ}{\rm C}$, promoting the growth of crystals at the bottom of the vial.

For single crystals of \ce{(PEA)2PbI4}, 223.2 mg of PbO (Sigma Aldrich) is dissolved in 2 ml of aqueous HI solution with the addition of 170 $\mu$L of 50\% aqueous \ce{H3PO2} (Sigma Aldrich). Separately, 92.4 $\mu$L
of phenethylamine (Sigma Aldrich) are neutralized in 1 ml of HI (57\% wt), yielding a white precipitate which re-dissolves upon heating. The latter solution is added to the PbO solution, and the mixture is stirred for 10 min at $150\,^{\circ}{\rm C}$ on a hotplate. Subsequently, the solution is left to cool down at room temperature. 24 h later, the grown orange crystals are collected via filtration and dried under vacuum at $100\,^{\circ}{\rm C}$.

\subsection{Raman measurements}
The Raman measurements were taken at $730$\,nm with an excitation power of $100$\,mW using a continuous wave Ti:sapphire laser from Spectra Physics (Matisse TS) in a near backscattering configuration. The beam was focused using a convex lens and the spot size was about 100\,$\mu$m. The spectra were detected by a Princeton Instruments CCD which is cooled by liquid nitrogen in conjunction with a double spectrometer Jobin-Yvon U1000. Everything was taken by a computer using an Arduino controlled by a Python program. The samples were always under vacuum inside the cryostat during the measurements. All the measurements were taken under vacuum using a cryostat with a Pfeiffer Vacuum HiCube turbo pump. The temperature was controlled with a CTI-CRYOGENICS compressor with a temperature controller Lakeshore $330$.

\subsection{Density functional theory calculations}
\label{dft_calculations}

The calculations have been performed by adopting periodic boundary conditions and localized atomic basis set as implemented in the CRYSTAL17 program \cite{dovesi_quantum-mechanical_2018}. The computational set-up consists of double split quality basis sets which include polarization, along with the PBE functional for the description of the exchange-correlation \cite{perdew_generalized_1996}. An automatic 4x4x1 sampling  of  the  first  Brillouin  zone  was  selected \cite{monkhorst_special_1976}, where  the  less  dense  sampling  is  related to  the  direction  associated  to  the  inorganic-sheet  stacking,  in  the  reciprocal  lattice.   The Grimme-D2 approach was included, to improve the description of the atomic forces between
the organic cations.  The SCF accuracy has been increased to 10$^{10}$ Hartree, to obtain accurate interatomic forces.  This computational set-up has been already tested for the parental CH$_{3}$NH$_{3}$PbI$_{3}$ perovskite  in  Ref.\cite{ivanovska_vibrational_2016} and  resulted  in  DFT  vibrational  spectra  in  excellent agreement with the experimental data available.

\begin{acknowledgement}


A.R.S.K.\ acknowledges funding from EU Horizon 2020 via a Marie Sklodowska Curie Fellowship (Global) (Project No.\ 705874). C.S.\ acknowledges partial support by the National Science Foundation (Award 1838276), the School of Chemistry and Biochemistry, and the College of Science of Georgia Institute of Technology. S.R.\ and R.L.\ acknowledge support from the Natural Science and Engineering Research Council of Canada. The work at Mons was supported by the Interuniversity Attraction Pole program of the Belgian Federal Science Policy Office (PAI 6/27) and FNRS-F.R.S. Computational resources have been provided by the Consortium des \'Equipements de Calcul Intensif (C\'ECI), funded by the Fonds de la Recherche Scientifique de Belgique (F.R.S.-FNRS) under Grant No.\ 2.5020.11. D.B. is a FNRS Research Director.

\end{acknowledgement}





\providecommand{\latin}[1]{#1}
\makeatletter
\providecommand{\doi}
  {\begingroup\let\do\@makeother\dospecials
  \catcode`\{=1 \catcode`\}=2\doi@aux}
\providecommand{\doi@aux}[1]{\endgroup\texttt{#1}}
\makeatother
\providecommand*\mcitethebibliography{\thebibliography}
\csname @ifundefined\endcsname{endmcitethebibliography}
  {\let\endmcitethebibliography\endthebibliography}{}

\end{document}